\documentclass[12pt,preprint]{aastex}
\usepackage{amsmath}
\usepackage[]{changes}
 \colorlet{Changes@Color}{red}
\setremarkmarkup{\footnote{#1\textcolor{blue#1}{#2}}}

\usepackage{color}           
 \definecolor{DarkGreen}{rgb}{0.0,0.45,0.0}  

\begin{document}

\title{High-Resolution Observations of Flares in an Arch Filament System}

\author{Yingna Su\altaffilmark{1,2}, Rui Liu\altaffilmark{3,4}, Shangwei Li\altaffilmark{1,5}, Wenda Cao\altaffilmark{6,7}, Kwangsu Ahn\altaffilmark{6}, Haisheng Ji\altaffilmark{1,2}}
\altaffiltext{1}{Key Laboratory of DMSA, Purple Mountain Observatory, Chinese Academy of Sciences, Nanjing 210008, China}  
\altaffiltext{2}{School of Astronomy and Space Science, University of Science and Technology of China, Hefei, Anhui 230026, Chin}
\altaffiltext{3}{CAS Key Laboratory of Geospace Environment, Department of Geophysics and Planetary Sciences, University
of Science and Technology of China, Hefei 230026, China}  
\altaffiltext{4}{Collaborative Innovation Center of Astronautical Science and Technology, Hefei 230026, China} 
\altaffiltext{5}{University of CAS, Beijing 100049, China}
\altaffiltext{6}{Big Bear Solar Observatory, New Jersey Institute of Technology, 40386 North Shore Lane, Big Bear City, California 92314-9672, USA}  
\altaffiltext{7}{Center for Solar-Terrestrial Research, New Jersey Institute of Technology, University Heights, Newark, New Jersey 07102-1982, USA}
\email{ynsu@pmo.ac.cn}
\email{rliu@ustc.edu.cn}

\begin{abstract}

We study five sequential solar flares (SOL2015-08-07) occurring in Active Region 12396 observed with the Goode Solar Telescope (GST) at the BBSO, complemented by IRIS and SDO observations. The main flaring region is an arch filament system (AFS) consisting of multiple bundles of dark filament threads enclosed by semi-circular flare ribbons. We study the magnetic configuration and evolution of the active region by constructing coronal magnetic field models based on SDO/HMI magnetograms using two independent methods, i.e., the nonlinear force-free field (NLFFF) extrapolation and the flux rope insertion method. The models consist of multiple flux ropes with mixed signs of helicity, i.e., positive (negative) in the northern (southern) region, which is consistent with the GST observations of multiple filament bundles. The footprints of quasi-separatrix layers (QSLs) derived from the extrapolated NLFFF compare favorably with the observed flare ribbons. An interesting double-ribbon fine structure located at the east border of the AFS is consistent with the fine structure of the QSL's footprint. Moreover, magnetic field lines traced along the semi-circular footprint of a dome-like QSL surrounding the AFS are connected to the regions of significant helicity and Poynting flux injection. The maps of magnetic twist show that positive twist became dominant as time progressed, which is consistent with the injection of positive helicity before the flares. We hence conclude that these circular shaped flares are caused by 3D magnetic reconnection at the QSLs associated with the AFS possessing mixed signs of helicity. 

\end{abstract}

\clearpage

\keywords{Sun: corona ---Sun: Chromosphere ---Sun: evolution --- Sun: flares --- Sun: filaments, prominences --- Sun: magnetic fields}

\section{INTRODUCTION}

The emergence of a new active region in chromosphere is generally recognized by a tiny bright plage and an H$\alpha$ arch filament system (AFS) \citep{Waldmeier1937, Bruzek1967, Weart1969, Zwaan1985}, which consists of a bundle of dark arches crossing the magnetic polarity inversion line (PIL) and connecting the regions of the innermost spots of opposite polarity. An AFS is typically visible for several days \citep{Bruzek1967}, while individual arch filaments only exist for a few tens of minutes. The arches in the AFS show a rising velocity of 10--15 km s$^{-1}$ at their tops and downflows of 10--50 km s$^{-1}$ at their endpoints, and have lengths of (1--3)$\times$10$^{4}$ km and a lifetime of the order of tens of minutes \citep[e.g.,][]{Shibata1989, Yoshimura1999}. The formation mechanism of AFS is well explained by the `leaky bucket' model \citep{Schmieder1991, Mein1996}. As new flux emerges, condensed material inside magnetic loops drains along both legs. With the material draining, the H$\alpha$ dark features of AFS get empty in a few minutes but new loops with condensed dark materials are formed below and the process can last for several hours, even a day. With the formation of new arch filaments, the old loops expand and can reach coronal heights \citep{Spadaro2004}. On the other hand,  \citet{Ma2015} found that the appearance and evolution of an AFS near the sunspot seems to be controlled by the moving magnetic features emanating from the penumbra. The AFS is different from the so-called PIL filament formed in a filament channel, which is defined as a region in the chromosphere surrounding a PIL where the chromospheric H$\alpha$ fibrils are aligned with the PIL \citep{Foukal1971, Gaizauskas1998}.

The linear force-free field extrapolations by \citet{Malherbe1998} show that AR NOAA 7785, in spite of having roughly a global potential configuration, consists of two systems of arch filaments. Moreover, these two systems are best fitted with two sheared structures of opposite $\alpha$ values of $\pm$0.1 Mm$^{-1}$. Using a magnetohydrostatic approach, \citet{Mandrini2002} analyzed the topology of Active Region (AR) NOAA 7968 from June 6 to June 9, and a surge \citep{Roy1973, Schmieder1995} occurred on June 9, 1996. They found that some of the arches of the AFS and the surge were associated with field lines having dips tangent to the photosphere (the so called ``bald patches''; BPs).  The observed evolution of the AFS and the surge is consistent with the expected results of 3D magnetic reconnection occurring in this magnetic topology.

Magnetic reconnections between newly emerging magnetic fields and pre-existing ambient fields can release a lot of energy and cause many eruptive activities, such as EUV brightenings or even flares \citep{Longcope2005, Zuccarello2008, Tarr2014}. \citet{Zuccarello2008} investigated the dynamics and the magnetic configuration of an AFS hosting a C-class flare, using a constant-$\alpha$ force-free magnetic field. They found that the interaction between new and pre-existing field lines, characterized by a small relative inclination, might have caused a weak reconnection process and given rise to the flare.

In this paper we study high resolution observations of the fine structure of an AFS as well as the associated flares during the emerging phase of the NOAA AR 12396 in order to understand the magnetic structure of the AFS as well as the initiation mechanisms of the flares within. We then compare them with extrapolated nonlinear force-free (NLFFF) fields, and investigate their evolution in terms of the topological changes in the magnetic field.

\section{Observations}

\subsection{Instrumentation and Data Set}

 On 2015  August 7, we carried out high-resolution observations of four flares occurred in NOAA AR 12396 with the 1.6-meter Goode Solar Telescope \citep[GST,][]{Cao2010} in the Big Bear Solar Observatory (BBSO). GST combines a high-order adaptive optics system using 308 sub-apertures and the post-facto speckle image reconstruction techniques \citep{Woger2008} to achieve diffraction-limited imaging of the solar atmosphere. The H$\alpha$ data are taken by the Visible Imaging Spectrometer \citep[VIS,][]{Cao2010}, which is a Fabry-P\'{e}rot filter-based system that can scan in the wavelength range of 5,500--7,000~{{\AA}}. The broadband TiO and H$\alpha$ images have a pixel scale of 0$\arcsec$.034 and 0$\arcsec$.029, respectively,  and the data are taken with a cadence of 15 s and 35 s, respectively. For this observation run, nine points are scanned at the H$\alpha$ line center and the line wings $\pm$0.4, $\pm$0.6, $\pm$0.8, $\pm$1.0~{{\AA}}. The TiO and H$\alpha$ images are co-aligned by matching sunspots and plage areas. The seeing condition at BBSO is mostly fair to good during this observing run. The images with bad seeings have been manually removed. About 56\% of the data after speckle reconstruction are used in the study. Corresponding ultraviolet and extreme ultraviolet observations are provided by AIA \citep{Lemen2012} onboard SDO. For the analysis of line of sight (LOS) photospheric magnetic fields, we use observations from HMI \citep{Schou2012} on board SDO with a cadence of 45 s (jsoc hmi.M$\_$45s series) and 1$\arcsec$/pixel spatial resolution. The HMI vector magnetograms have a cadence of 12 minutes. The vector field data taken by the Near InfraRed Imaging Spectropolarimeter \citep[NIRIS,][]{Cao2012, Wang2017} at the 1.56-$\mu$m Fe {\sc i} line are also presented. The pixel size is 0$\arcsec$.08, and the time cadence is 67 seconds. The fifth flare on 2015 Aug 8 is observed by the Interface Region Imaging Spectrograph \citep[IRIS,][]{Depontieu2014} at 1330, 1400, and 2796~{{\AA}} with large coarse 8-step raster mode at a cadence of 34 s.

The coalignment between GST data and SDO data is processed in the following three steps. At first, one GST image in the H$\alpha$ line wing (e.g., +1.0~\AA) is overlaid on the closest in time HMI continuum image. The coalignment is achieved by comparing distinct features such us sunspots in the two images by eyes. Thus we can obtain the rotation angle and coordinate information of the GST image, then we derotate the GST image by taking the offset into account. The second step is the alignment of the GST images in time series. This coalignment is operated by maximizing the cross-correlation using fast fourier transform(FFT) method. We align the images frame by frame, and the first reference frame is the image we obtained during the first coalignment. The third step is the coalignment of the GST images at other passbands. Given that the observation time interval in different bands is relatively short (about 2 seconds), we assume that the rotation angle and coordinate information in all of the H$\alpha$ line center and line wings are the same. Similar to the H$\alpha$ images, the GST TiO images are also aligned with the corresponding HMI continuum images.

\subsection{Dynamics of Filament Threads}

\begin{figure} 
\begin{center}
\epsscale{1} \plotone{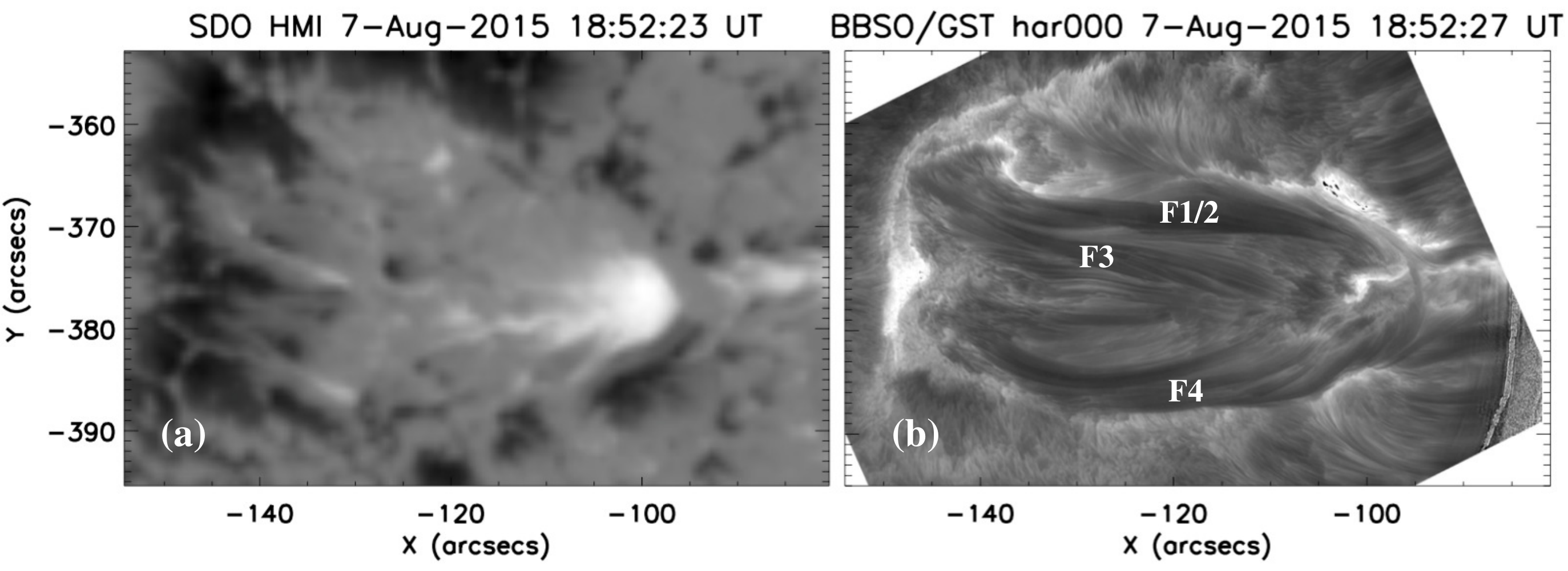}     
\end{center}
\caption{(a) Line of sight photospheric magnetogram acquired by SDO/HMI on 2015 Aug 7 showing the magnetic configuration of NOAA 12396; (b) target region observed at H$\alpha$ line center by GST. Note that the dark area within the bright ribbons in the GST H$\alpha$ image is due to CCD saturations.}
\label{fig:filament-preview}
\end{figure}

The active region of interest (AR 12396) is composed of a leading sunspot group of positive polarity encircled by a chain of following sunspots of negative polarity (see Figure \ref{fig:filament-preview}a, showing the relevant magnetogram). The evolution of the active region over this time interval is featured by both flux emergence and shearing flows. Our analysis suggests that this active region is in its emerging phase. 
Figure \ref{fig:filament-preview}b reports a H$\alpha$ line center image acquired by GST showing that the main flaring region encloses bundles of dark filament threads crossing the magnetic polarity inversion line and connecting spots of opposite polarity, conforming to an arch filament system reported in the literature. The H$\alpha$ line center image also shows that the structure of this system is very complex with several isolated filament bundles darker than the others, namely F1/2, F3, F4.

\begin{figure} 
\begin{center}
\epsscale{1} \plotone{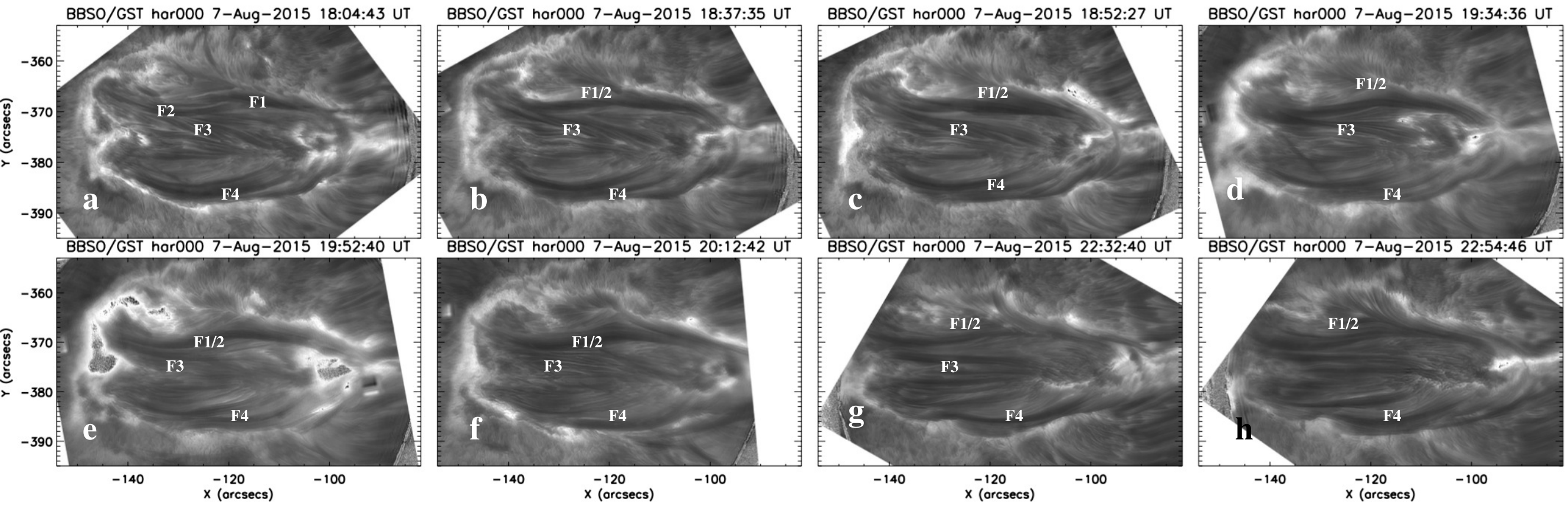}     
\end{center}
\caption{Evolution of filament threads at H$\alpha$ line center observed by GST on 2015 Aug 7. The rectangle artifacts at the west end of the AFS in panel (e) sometimes appear due to speckle reconstructions when it is performed to low-contrast data taken under a fair/poor seeing condition. Corresponding animation video 1 can be found online. }
\label{fig:gst-ha}
\end{figure}

A series of H$\alpha$ line center images taken by GST before and during the four flares on August 7 are shown in Figure~\ref{fig:gst-ha} and corresponding online animation named video 1. The time cadence of the video is not stable, since some of the images with bad quality are manually removed. This region is composed of a series of dark filament threads surrounded by semi-circular bright ribbons. The filament threads are very dynamic and have complex structure. At 18:04 UT, the northern region contains mainly three bundles of darker filament threads F1, F2, and F3 (Figure~\ref{fig:gst-ha}a), and several filament threads showing similar behavior in the southern region named F4. The two J-shaped filament threads F1 and F2 gradually merge with each other and form a longer filament F1/2.  F1/2 appears to have a straightened `S' shape (Figures \ref{fig:gst-ha}b--2c), which becomes more curved later (Figure \ref{fig:gst-ha}d). This merging process is also presented by \citet{Mandrini2002}, who associated this phenomenon with the split BP topology. The comparison between the top and the bottom panels of Figure \ref{fig:gst-ha} shows that filament F3 appears to evolve from an inclined twisted structure to a more straightened structure like the traditional arch filaments (also see video 1). Part of the filament threads F4 in the southern region rises up and disappears. Detailed descriptions can be found in \S\ref{sec:flare}.

\subsection{Dynamics of Flares} \label{sec:flare}

\begin{figure} 
\begin{center}
\epsscale{0.9} \plotone{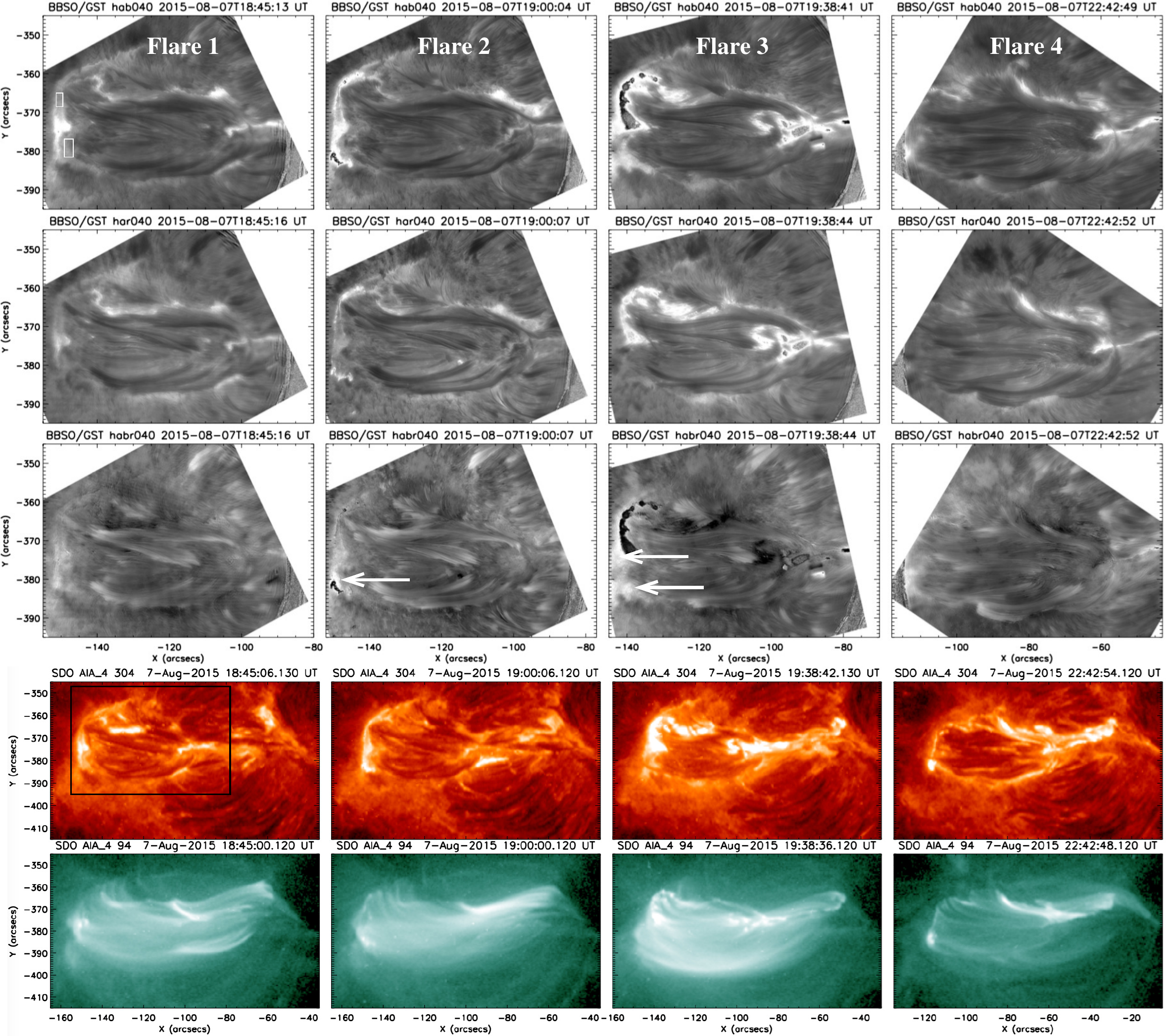}     
\end{center}
\caption{Overview of four flares observed by GST and AIA on 2015 Aug 7. The first and second rows show images taken at H$\alpha$ blue and red wings, the corresponding dopplergrams are presented in the third row, and AIA images in 304~{\AA}~and 94~{\AA}~are displayed on the last two rows, respectively. Note that in dopplergrams brighter (darker) regions indicate upflows (downflows). Corresponding animation video 2 can be found online.}
\label{fig:flares}
\end{figure}

The five flares of interest are listed in Table 1, identified using the IAU SOL target naming convention by \citet{Leibacher2010}. Figure \ref{fig:flares} presents an overview of the four flares taken by GST and AIA, and corresponding animations (video 2) can be found online. The GST H$\alpha$ observations show that bright flare ribbons are mostly surrounding the AFS.  One can see that the flare ribbons in the blue wing (H$\alpha$-0.4~{\AA}; first row) are brighter than those in the red wing (H$\alpha$+0.4~{\AA}; second row) for all four flares. Note that the VIS has been carefully calibrated to find the H$\alpha$ line center ahead of normal operation each day, and also the tuning of VIS is very precise. However, the profile of VIS prefilter is no longer symmetric. Therefore, we choose two locations without flare-resulted emission as background at 18:45 UT (marked with two small white boxes in Figure \ref{fig:flares}) in order to obtain the ratio between the emission in blue and red wings. We then obtain dopplergrams after the line asymmetry correction. The resulted dopplergrams (3rd row of Figure \ref{fig:flares}, brighter/darker regions refer to upflow/downlow) show that some locations with strongest emission are likely to be real blue shift (marked with white arrows).  From the semiempirical point of view, it is shown that chromospheric condensations can be responsible for  the H$\alpha$ blue line asymmetry with a central reversal \citep{Gan1993}. The emission asymmetry at other locations is mainly due to the asymmetry of VIS prefilter. We also find that the footpoints of the dark filament threads are darker in red wing (red shifted), while the top of the filament threads appear to be darker in blue wing (blue shifted). The dopplergrams show the opposite, i.e., brighter in footpoints and darker on top, because the filament threads are observed due to absorption. Corresponding AIA 304~{\AA}~images with a larger field of view (FOV) present the full picture of the bright ribbon to the west of the AFS. Multiple flare loop systems are identified in AIA 94~{\AA}~images. We find that the bright ribbons in the four flares are mostly located at the northeast and west border of the AFS, while the brightened area in Flare 3 appears to be the largest. The flare ribbons at the east border of the AFS consist of two closely spaced ribbons (referred to as a \emph{double ribbon} hereafter), which is revealed in Flares 1--2 by high-resolution GST H$\alpha$ images. This structure is unfortunately out of the GST FOV in Flare 4. Below we will elaborate on three representative flares.  

\subsubsection{Flare 3 with Untwist Motion}
\begin{figure} 
\begin{center}
\epsscale{0.9} \plotone{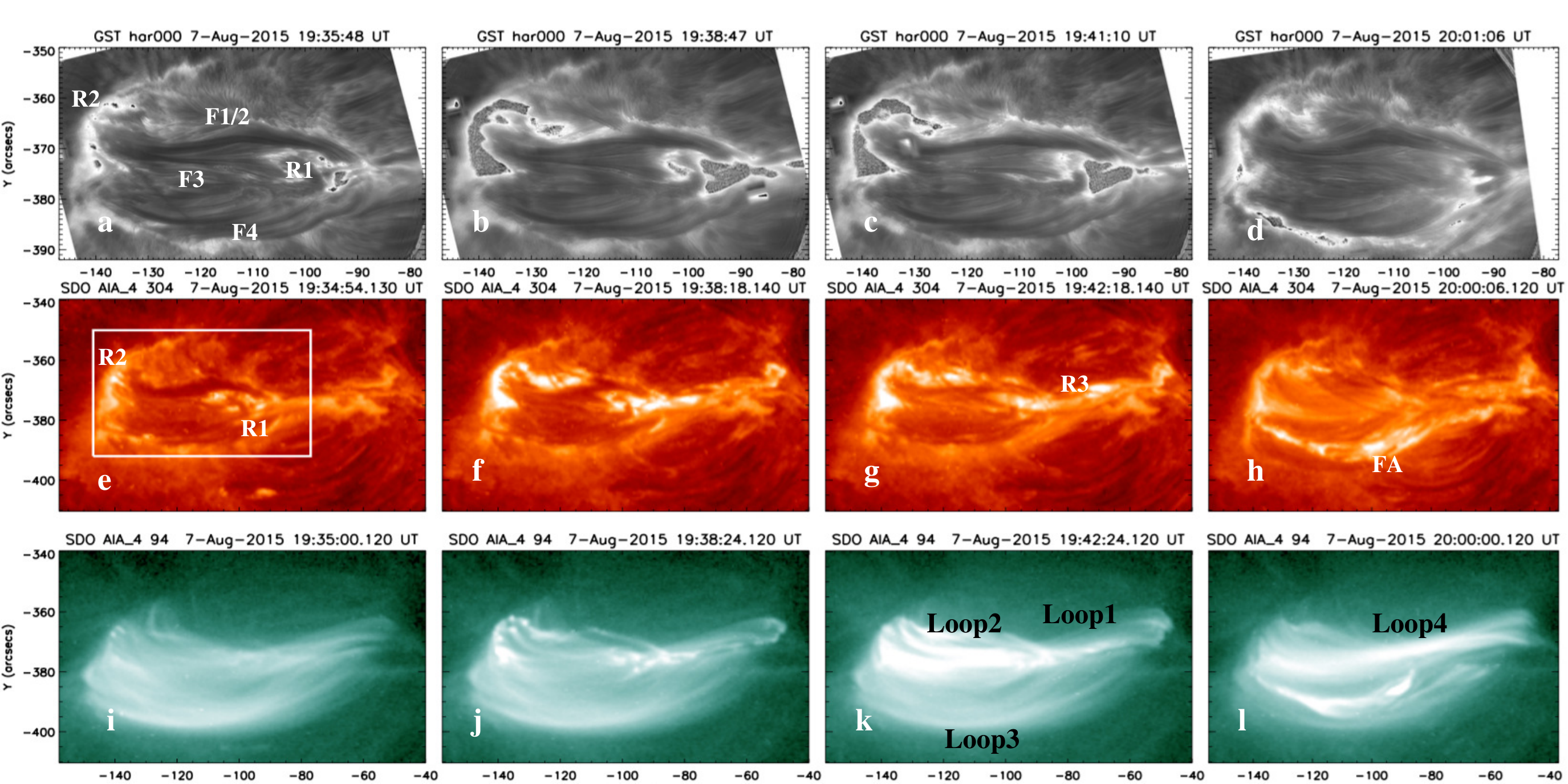}     
\end{center}
\caption{GST H$\alpha$ line center (first row) and AIA images in 304~{\AA}~(second row) and 94~{\AA}~(third row) of Flare 3 at 19:35 UT (first column), 19:38 UT (second column), 19:41 UT (third column), and 20:01 UT (fourth column) on 2015 Aug 7, respectively.}
\label{fig:flare3}
\end{figure}

The morphological evolution of Flare 3 taken by GST and AIA are presented in Figure \ref{fig:flare3}.  At 19:34 UT, Flare 3 begins with brightenings R1 and R2  (Figure \ref{fig:flare3}e), which are located near the footpoints of filament F3 in high-resolution H$\alpha$ images. These brightenings  then extend to the footpoints of filament F1/2 with time. A dark S-shaped filament is observed in 304~{\AA} corresponding to filament threads F1/2 in H$\alpha$. Filament F3 begins to rise up around 19:39 UT and appears to display a transient counterclockwise rotation if looking from the west (Figures \ref{fig:flare3}b--\ref{fig:flare3}c and online animation videos 1 and 2). During this process filament F3 evolves from an inclined twisted dark structure to a more standard arch filament structure with reduced darkness. This small untwisting motion is only marginally visible in the AIA observations due to limited resolution. Around 19:43 UT, filament bundles F4 in the southern region rise up and activate, during which the southern border of the AFS is brightened (the fourth column). The AIA images in 94~{\AA}~(Figures \ref{fig:flare3}k and \ref{fig:flare3}l) show that the post-flare loops consist of at least four loop systems.

\subsubsection{Flares with Double Ribbons}

\begin{figure} 
\begin{center}
\epsscale{0.9} \plotone{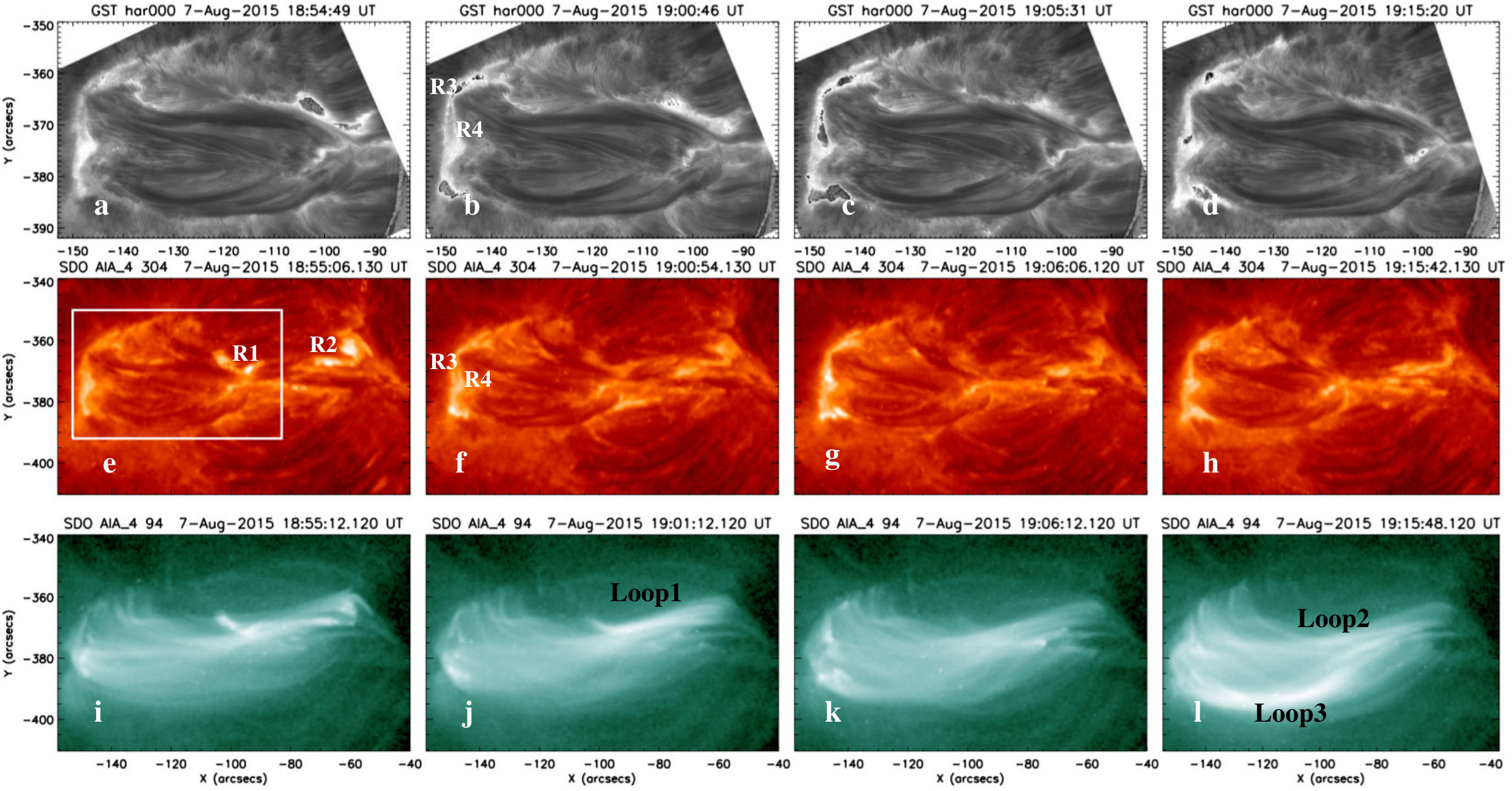}     
\end{center}
\caption{GST H$\alpha$ line center (first row) and AIA images in 304~{\AA}~(middle row) and 94~{\AA}~(bottom row) of Flare 2 at 18:54 UT, 19:00 UT, 19:05 UT, and 19:15 UT on 2015 Aug 7, respectively. }
\label{fig:flare2}
\end{figure}

Figure \ref{fig:flare2} presents the morphological evolution of Flare 2 observed by GST and AIA. Flare 2 begins with brightenings R1 and R2 at the west border and a small bright point at the east border of the AFS around 18:52 UT (left column). An elongated ribbon R3 starts to brighten up at 18:57 UT, which is followed by the brightening of ribbon R4 at 19:00 UT. The most interesting feature of this flare is that the eastern flare ribbon is composed of a pair of closely spaced ribbons (double-ribbon), i.e., R3 and R4, which is clearly visible in the high-resolution H$\alpha$ observations by GST. The double-ribbon structure can also be identified in the corresponding AIA 304~{\AA}~images, but not as clear. The corresponding AIA 94~{\AA}~images show that the post-flare loops are composed of at least three loop systems. 

\begin{figure} 
\begin{center}
\epsscale{0.9} \plotone{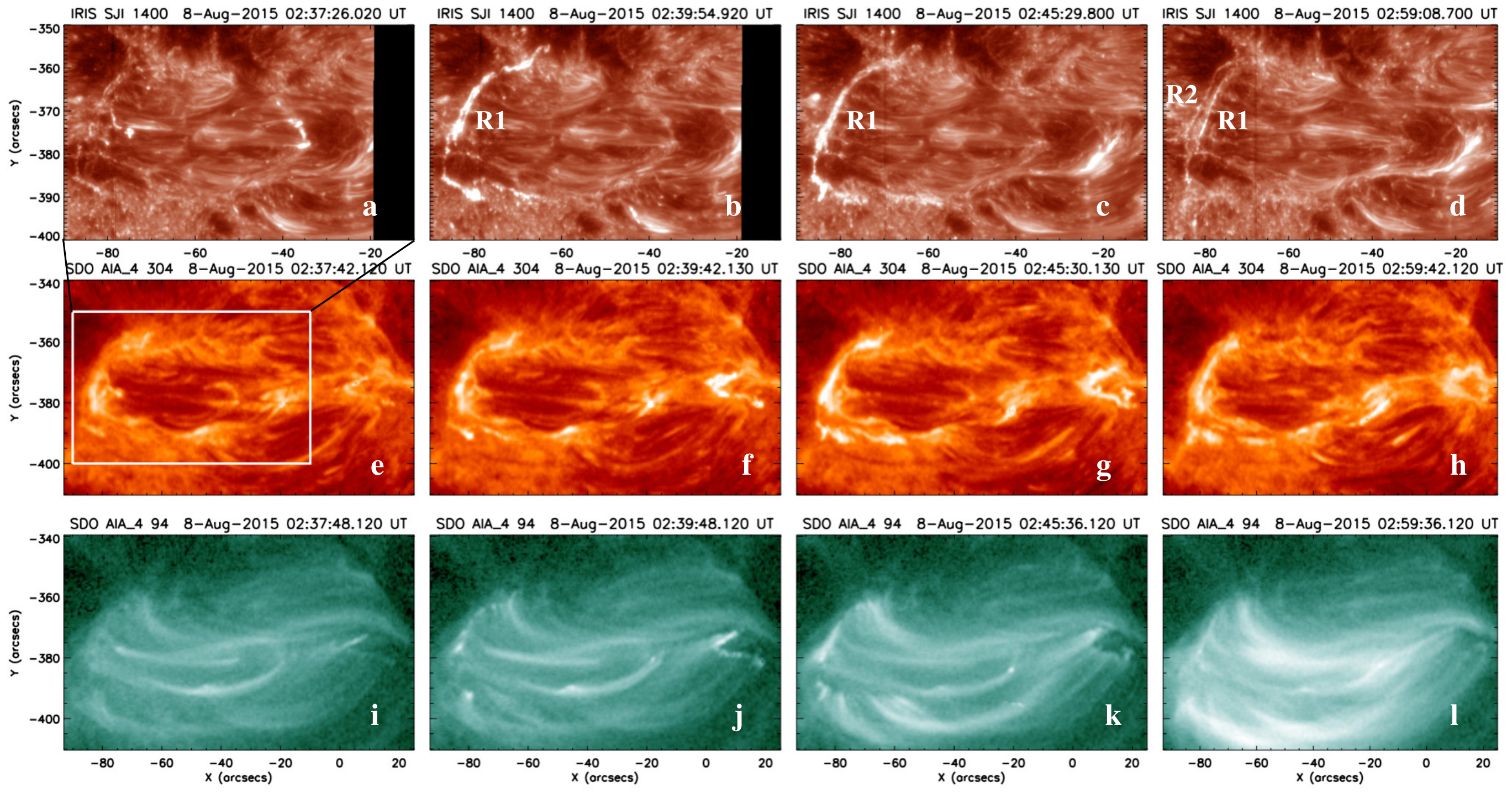}     
\end{center}
\caption{IRIS images in 1400~{\AA}~(first row) and AIA images in 304~{\AA}~(second row) and 94~{\AA}~(third row) of Flare 5 at 02:37 UT (first column), 02:39 UT (second column), 02:45 UT (third column), and 02:59 UT (fourth column) on 2015 Aug 8, respectively.}
\label{fig:flare5}
\end{figure}

The aforementioned double-ribbon structure is also visible in Flares 1 and 5 as presented in Figures \ref{fig:flares} and \ref{fig:flare5}. Flare 5 is not observed by GST but IRIS.  IRIS 1400~{\AA} slit-jaw images show that Flare 5 begins with brightenings in ribbon R1 at the east border of the AFS at 02:38 UT on August 8. R1 appears to also consist of a fine double-ribbon structure before it reaches saturation as shown in Figure \ref{fig:flare5}c. With the fading of R1, ribbon R2 brightens up around 02:54 UT, followed by the re-brightening of ribbon R1 at 02:56 UT. Similar to Flare 2, R1 and R2 are a pairs of closely spaced ribbons. The evolution of the double-ribbon (R1 and R2) structure can be clearly identified in the high-resolution IRIS 1400~{\AA} images, but barely visible in the corresponding AIA 304~{\AA}~images (2nd row in Figure~\ref{fig:flare5}). The multiple post-flare loop systems are shown in the AIA 94~{\AA} images (3rd row in Figure~\ref{fig:flare5}).
 
\subsection{Magnetic Configuration and Evolution of AR 12396}

\begin{figure} 
\begin{center}
\epsscale{0.8} \plotone{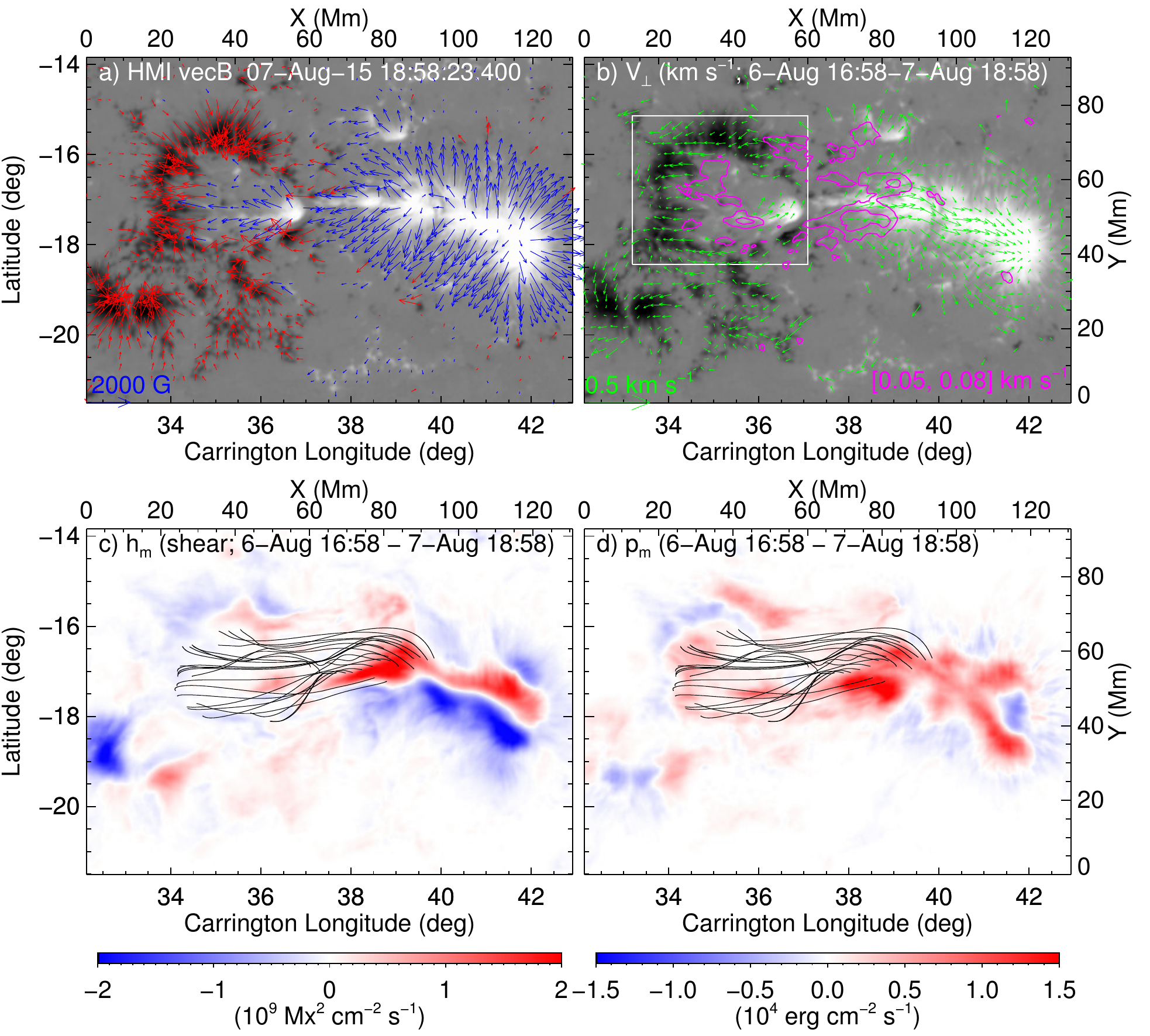}     
\end{center}
\caption{SDO/HMI observations of magnetic field evolution before the flares of interest. Panels (a) and (b) present vertical components of magnetic fields overlaid with horizontal magnetic vectors at 18:58 UT on Aug 7, and with photospheric flows perpendicular to magnetic field, which are averaged from 16:58 UT on Aug 6 to 18:58 UT on Aug 7 2015. The green arrows refer to the transverse velocity vectors. The magenta contours in (b) indicate the normal velocity ($V_{\perp n}$) at 0.05 and 0.08 km s $^{-1}$ (upflows). Panels (c) and (d) show the map of helicity flux density (shear term) and of Poynting flux density, respectively, averaged over the time interval from 16:58 UT on Aug 6 to 18:58 UT on Aug 7 2015. The two maps are superimposed with field lines traced from the semi-circular high-$Q$ line (blue) that surrounds the dark filament system (see Figure~\ref{fig:topology}c2).}
\label{fig:bfield}
\end{figure}

\begin{figure} 
\begin{center}
\epsscale{0.8} \plotone{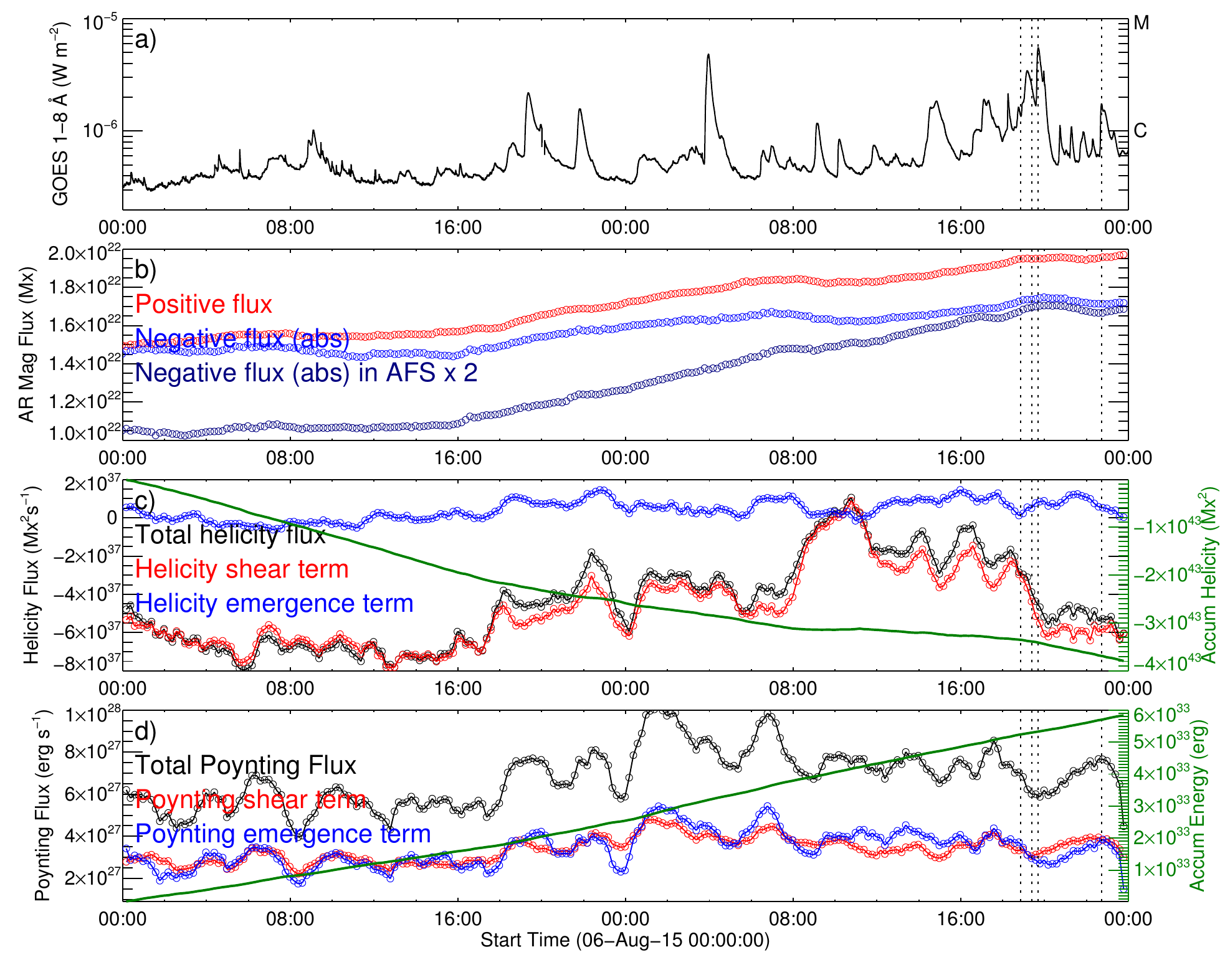}     
\end{center}
\caption{Temporal evolution of X-ray and various magnetic parameters in AR 12396 before and during the flares of interest. (a) GOES soft X-ray light curve in 1--8~{\AA}. The peak time of the four flares observed by BBSO/GST is marked by dotted vertical lines. (b) Temporal evolution of positive (red) and negative flux (blue) and twice of the negative flux in the dark filament system (dark blue) marked by a white box in Figure~\ref{fig:bfield}b. (c) Light curves of the shear (red) and emergence (blue) term of the relative helicity flux across the photospheric boundary, the sum of the two terms (total helicity flux; black), and the accumulative helicity (green; scaled by the y-axis on the right). (d) Temporal evolution of the shear (red) and emergence (blue) of the Poynting flux across the photospheric boundary, the sum of the two terms (total Poynting flux; black), and the accumulative magnetic energy in the corona (green; scaled by the y-axis on the right).}
\label{fig:ltc-ar} 
\end{figure}

In order to study how the flare energy is built up and released as well as the initiation mechanism, we study the magnetic configuration and evolution of AR 12396 before and during the flares. In particular, we study the temporal evolution of photospheric flows, magnetic flux, helicity flux, and Poynting flux. The magnetic configuration of AR 12396 is characterized by a leading sunspot group of positive polarity, which has a feather shape, and a chain of following sunspots of negative polarity, which is semicircularly aligned around the stem of the `white feather' (Figure~\ref{fig:bfield}a). We obtained the photospheric flow field by applying the Differential Affine Velocity Estimator for Vector Magnetograms \citep[DAVE4VM;][]{Schuck2008} to the time-series of deprojected, registered vector magnetograms. The flow field is then subtracted by the field-aligned plasma flow $(\mathbf{V}\cdot\mathbf{B})\mathbf{B}/B^2$ to yield $\mathbf{V}_\perp$ that is perpendicular to magnetic field. The flow map shown in Figure~\ref{fig:bfield}b is averaged over the time interval from 16:58 UT on 2015 August 6 until 18:58 UT on August 7, the beginning of the sequential flares observed by GST (Table 1). In Figure~\ref{fig:bfield}b, green vectors refer to the transverse component $\mathbf{V}_{\perp t}$, while magenta contours indicate the upflow component $V_{\perp n}$ at 0.05 and 0.08 km s$^{-1}$. During this time interval, AR 12396 experienced flux emergence (Figure \ref{fig:ltc-ar}b) and an injection of positive helicity (Figure \ref{fig:ltc-ar}c), the latter of which, however, is dominated by the shear ($\mathbf{V}_{\perp t}$) rather than the emergence ($V_{\perp n}$) term of helicity \citep{Berger1984, Liu2012, Liu2014}. On the other hand,  both flux emergence ($V_{\perp n}$) and tangential motions ($\mathbf{V}_{\perp t}$) contribute to the Poynting flux across the photospheric boundary (Figure \ref{fig:ltc-ar}d). For detailed definition of the aforementioned terms, please refer to \citet{Liu2016}.  As revealed in the maps of flux density (Figure \ref{fig:bfield}c and \ref{fig:bfield}d), the stem of feather, corresponding to the stem of the tulip-shaped AFS, appears to be a place with significant injection of both positive helicity and Poynting fluxes.

\begin{figure} 
\begin{center}
\epsscale{0.8} \plotone{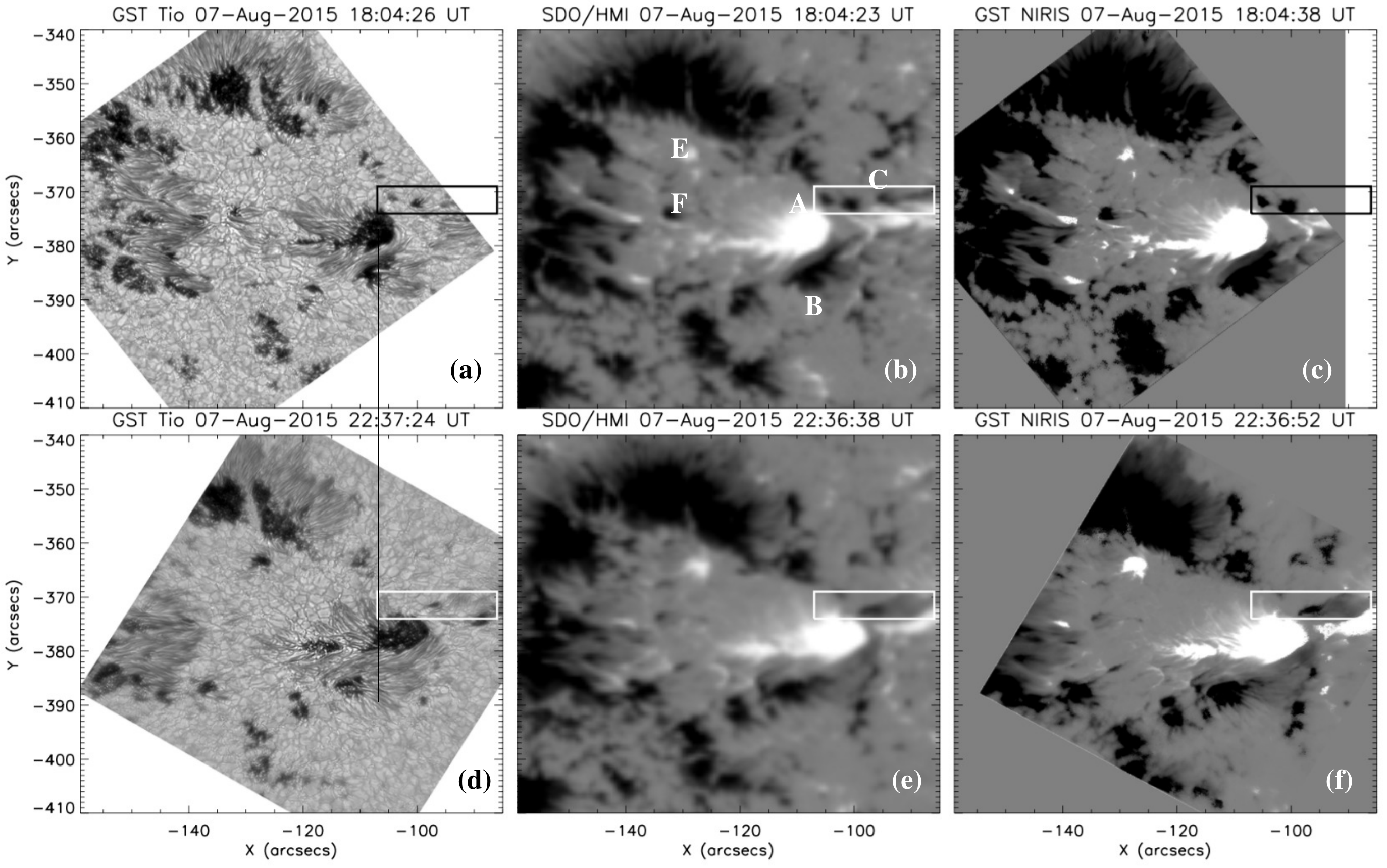}     
\end{center}
\caption{GST/TiO images (first column) and photospheric magnetograms observed by HMI (second column) and GST (third column) at 18:04 UT (top row) and 22:36 UT (bottom row). Significant flux cancellations occur in the region enclosed by black and white boxes.}
\label{fig:evo-bfield}
\end{figure}

We now get into details of the evolution of photospheric magnetic fields relevant to the four flares observed on 2015 August 7 (Figure \ref{fig:evo-bfield}). From left to right, columns show TiO images by GST, LOS photospheric magnetograms by SDO/HMI, and vertical component of photospheric vector magnetic fields by NIRIS. The NIRIS data undergo Stokes inversion using the Milne-Eddington technique, through which several key physical parameters (including total magnetic field, azimuth angle, inclination, and Doppler shift) can be extracted. For successful fittings with Milne-Eddington-simulated profiles, initial parameters are pre-calculated to be close to the observed Stokes profiles. The accuracy of the resulted vector field data is 10 G for the LOS component and 100 G for the transverse component. Comparing images in the top and bottom rows taken before Flare 1 (18:04 UT) and after Flare 4 (22:37UT), we find significant changes at several locations. Clear shear motions occurred in the two sunspots A and B during GST observing time interval on August 7, with A moving westward and B eastward, which can be identified from the dark reference line in Figures \ref{fig:evo-bfield}a and \ref{fig:evo-bfield}d. More highly sheared penumbra fibrils developed subsequently between A and B. Pore F moved eastward then gradually fragmented below the detectable resolution. Several small positive polarities gradually moved westward and joined Port E, which leads to the size increase of Pore E. Moreover, flux cancellations occurred in Box C (Figure \ref{fig:evo-bfield}b) during the time period of the four flares of interest. This can be identified by comparing the images in the top and bottom rows.

\section{Magnetic Field Modeling}

\subsection{NLFFF Extrapolations}

\begin{figure} 
\begin{center}
\epsscale{1} \plotone{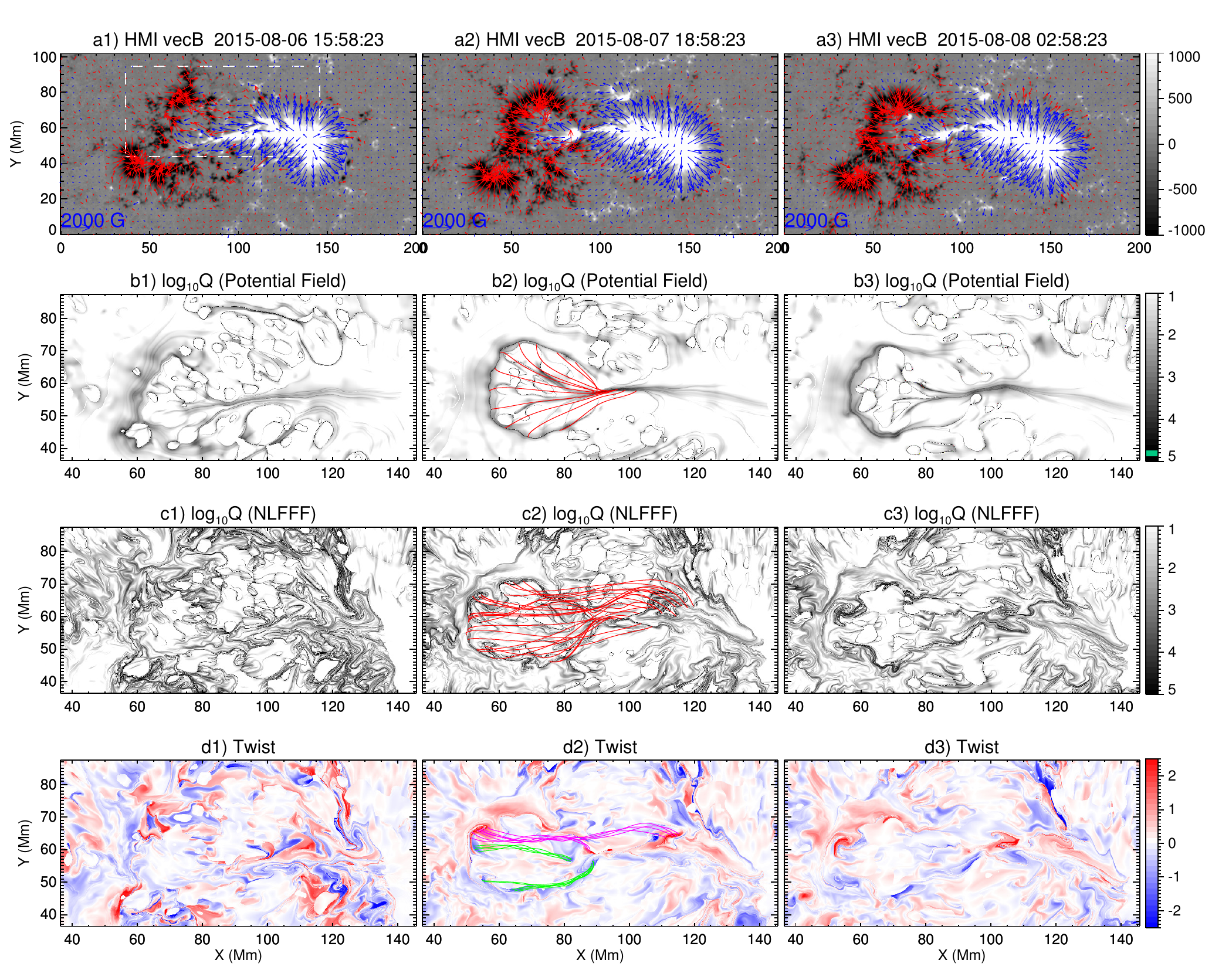}     
\end{center}
\caption{Evolution of AR 12396 from Aug 6 to Aug 8 in 2015. (a1--a3) HMI vector magnetograms. The rectangle in a1) indicates the FOV for the panels below. (b1--b3) Photospheric maps of $\mathrm{log}\,Q$ for the potential field extrapolation. (c1--c3) photospheric maps of $\mathrm{log}\,Q$ for the NLFFF extrapolation. (d1--d3) Photospheric maps of magnetic twist derived from the NLFFF (see \citet{Liu2016}). Panels (b2) and (c2) are superimposed with selected field lines traced from the semi-circular high-$Q$ line (blue) that surrounds the AFS of interest. Panel (d2) is superimposed with selected field lines with significant magnetic twist. Field lines of positive (negative) twist are indicated in magenta (green).}
\label{fig:topology}
\end{figure}

To understand the magnetic connectivities within the active region and their evolution, we use the code package developed by T. Wiegelmann, which utilizes the ``weighted optimization" method \citep{Wiegelmann2004,Wiegelmann2012} to build a NLFFF model to approximate the coronal field. To best suit the force-free condition, the vector magnetograms are ``pre-processed" \citep{Wiegelmann2006SoPh} before being taken as the photospheric boundary. Our calculation is performed within a box of $880 \times 328 \times 328$ uniformly spaced grid points, whose photospheric FOV is shown in the top row of Figure \ref{fig:bfield}. The potential field is calculated by a Fourier transformation method. We then calculate squashing factor $Q$ \citep{Titov2002} for both potential field (Figures~\ref{fig:topology}b1--b3) and NLFFF (Figures~\ref{fig:topology}c1--c3), and magnetic twist for NLFFF \citep[Figures~\ref{fig:topology}d1--d3; see][for methodology]{Liu2016} in a box region covering the rectangle in Figure~\ref{fig:topology}a1.

\begin{figure} 
\begin{center}
\epsscale{0.8} \plotone{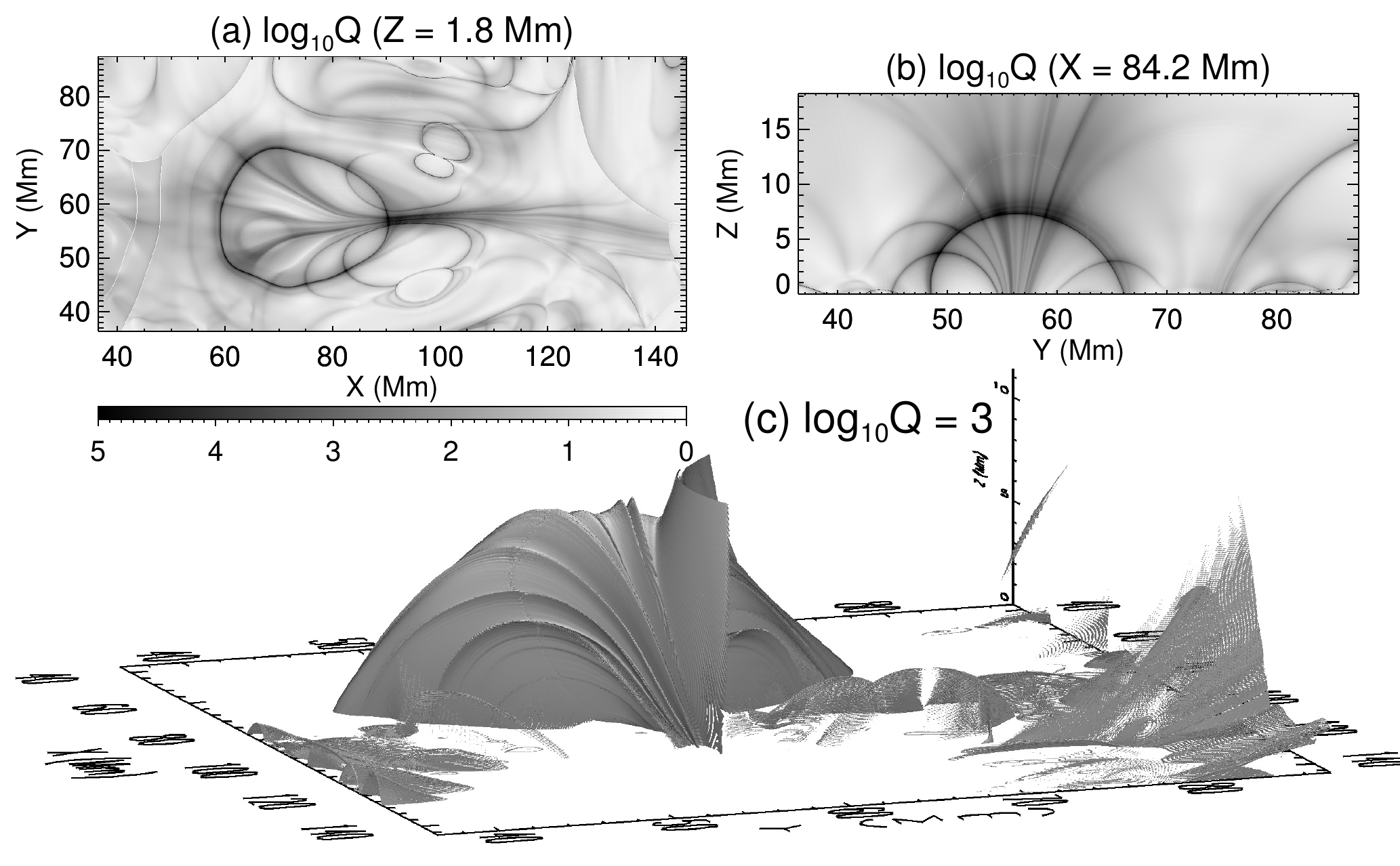}     
\end{center}
\caption{(a), (b), (c) 2D cuts in XY, YZ planes, and 3D high-Q surfaces based upon the potential field model at 18:58:23 UT on 2015 August 7, respectively.}
\label{fig:pot-q3d}
\end{figure}

The magnetic topology associated with the AFS is clearly indicated by the $Q$-factor derived from potential field. A high-$Q$ line associated with the semicircular chain of negative spots stands out in the $\mathrm{log}\,Q$ map on 2015 August 7 (Figures~\ref{fig:topology}b2--b3). This high-$Q$ line is the photospheric footprint of a dome-like quasi-separatrix layer (QSL), carving vertically by multiple high-$Q$ surfaces, whose footprints correspond to the high-$Q$ lines that are roughly east-west oriented. The intersections between these high-$Q$ surfaces and the dome are the preferential places for the formation of current sheets and subsequent dissipation through magnetic reconnection. The high-Q surfaces in 3D perspective is presented in Figure~\ref{fig:pot-q3d}. From this figure, one can clearly see how the multiple curved high-Q surfaces cut through a dome-shaped QSL, and these high-Q surfaces converge towards the west. The intersection of each high-Q surface and the dome-shaped QSL is a 3D curve, which is a quasi-separator by definition \citep[Section 6.9.5,][]{Priest2014}. In 2D cuts (XY and YZ planes), the intersection gives an X shape, which is usually considered as a 3D generalization of the X point in the 2D reconnection models. This is very different from the null-point topology. The 3D Q-factor is based upon the potential field model at 18:58:23 UT on 2015 August 7.

\begin{figure} 
\begin{center}
\epsscale{0.8} \plotone{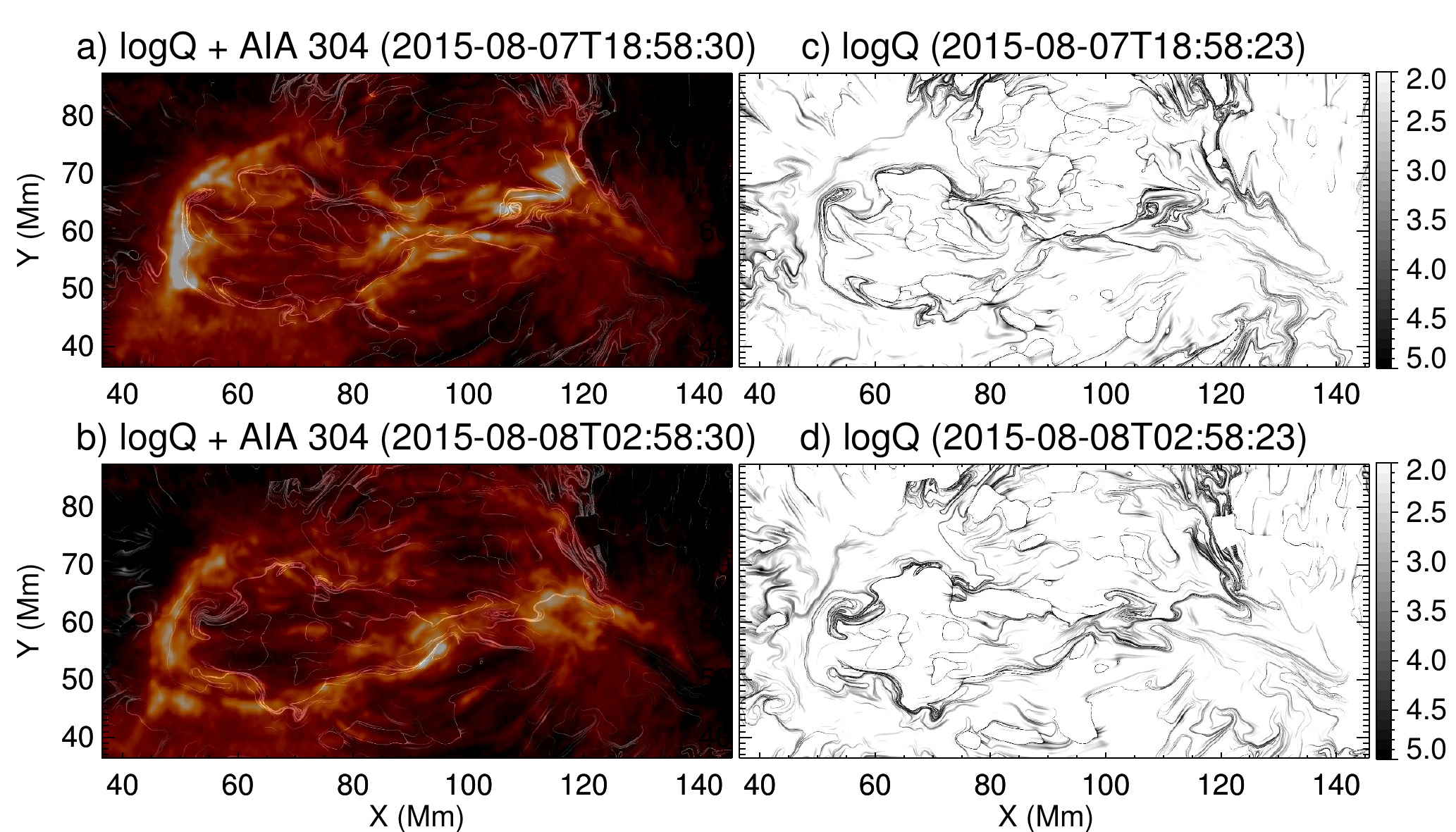}     
\end{center}
\caption{Comparison of $\log Q$ maps with observed double ribbons. Left column shows the blended images of AIA 304~{{\AA}} with corresponding maps of photospheric $\log Q$ for NLFFF extrapolations, the latter of which are displayed in the right column. The AIA images are projected with the cylindrical equal area (CEA) method to the same FOV of the $\log Q$ map.}
\label{fig:comp}
\end{figure}

Though the basic topology is retained, the magnetic connectivity in NLFFF is much more complicated than that in potential field, which is demonstrated by the $\mathrm{log}\,Q$ maps (Figures~\ref{fig:topology}c1--c3), as well as by the field lines traced from the semi-circular blue high-$Q$ line surrounding the flaring region (Figures~\ref{fig:topology}b2--c2). These field lines are apparently rooted in the regions of significant helicity and Poynting flux injection (Figures~\ref{fig:bfield}c--d).  The semi-circular high-Q line in the east together with the east-west oriented high-Q lines in the west (Figure~\ref{fig:topology}c2) compares favorably with the observed flare ribbons (Figure~\ref{fig:comp}), which is consistent with previous studies \citep[e.g.,][]{Su2009, Savcheva2015, Savcheva2016, Liu2016, Kang2017}. Despite a small displacement, the complex high-$Q$ structures also nicely match the double ribbon structure as shown in Figure~\ref{fig:comp}.  

Associated with the injection of positive helicity, positive twist becomes dominant in this region, at the beginning of the sequential flares (Figure~\ref{fig:topology}d2). We are able to identify three pertinent flux ropes by tracing field lines with significant twist numbers, including a rope with positive twist (magenta field lines) in the north, which apparently possesses two branches, and two ropes with negative twist (green field lines) in the south (Figure~\ref{fig:topology}d2). These flux ropes resemble the filament threads within the AFS. A clear decrease of negative twist of the northern flux rope represented by green field lines (marked with a black arrow) is identified by comparisons of twist map at 18:58 UT on Aug 7 and 02:58 UT on Aug 8. The twist number is defined as an integration of $J \cdot B / B^2$ along the field line. Please see Equation 16 in \citet{Berger2006}, and also see Appendix C in \citet{Liu2016} for its difference from the number of turns of a field line winding around a specified axis.

\subsection{Flux Rope Insertion Method}

In order to compare with the aforementioned NLFFF extrapolations' results, we construct a series of magnetic field models using the flux rope insertion method developed by \citet{vanBallegooijen2004}. A detailed description of the methodology can be found in the literature \citep{Bobra2008, Su2011, Su2012}. First, a potential field is computed from the high-resolution (HIRES) magnetogram embedded in a low-resolution global map. Then, by appropriate modifications of the vector potentials, a ``cavity" is created above a selected path, and a thin flux bundle representing the axial flux of the flux rope ($\Phi_{axi}$) is inserted into the cavity. Circular loops are added around the flux bundle to represent the poloidal flux of the flux rope (F$_{pol}$). The above field configuration is not in force-free equilibrium, so our next step is to make the field evolve toward a force-free state through magnetofrictional relaxation. This  is an iterative relaxation method \citep{Yang1986, vanBallegooijen2000} specifically designed for vector potentials.

\begin{figure} 
\begin{center}
\epsscale{1.0} \plotone{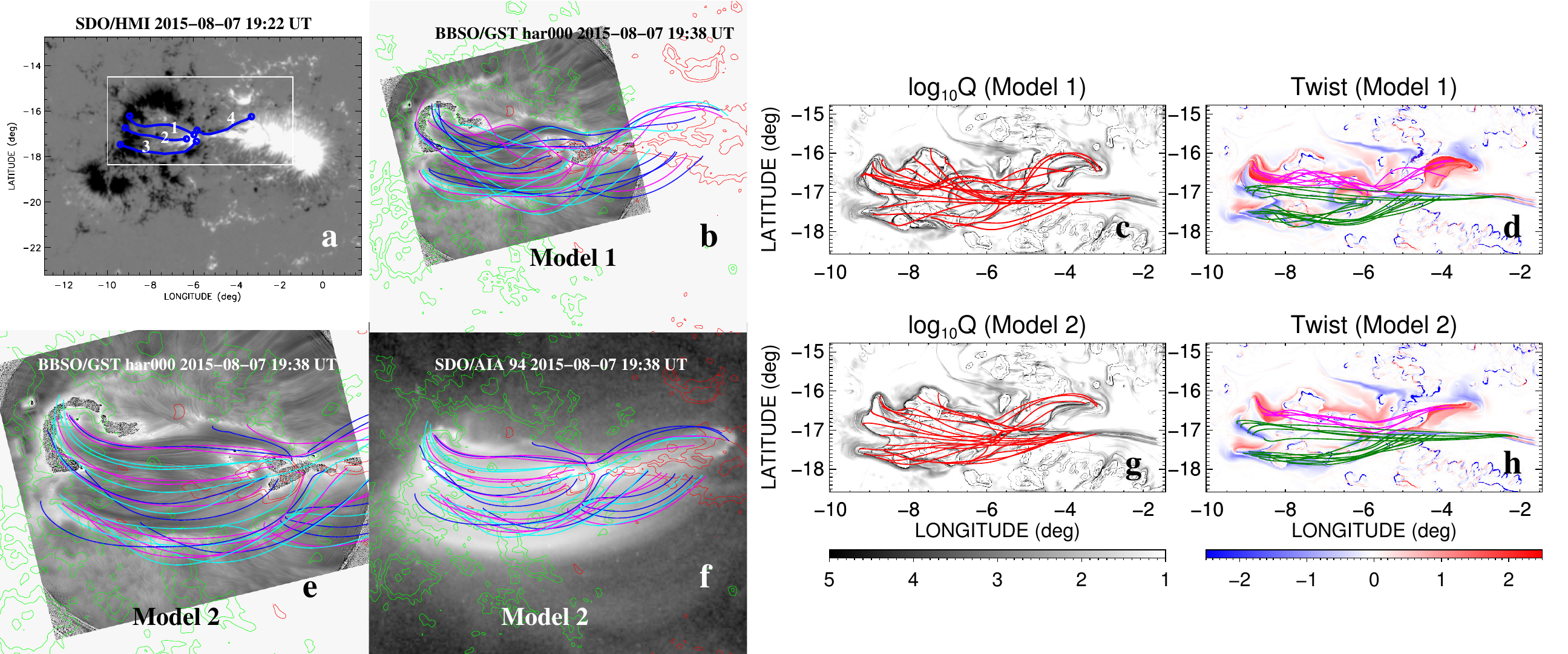}     
\end{center}
\caption{Flux rope insertion method. (a) Longitude--latitude map of the radial component of the magnetic field in the photosphere in the HIRES region of the model at 19:22 UT on 2015 Aug 7. The blue curves show the paths along which the flux ropes are inserted into the model.
(b) Selected magnetic field lines from Model 1 overlaid on H$\alpha$ line center image. The red and green contours refer to the positive and negative magnetic polarities observed by HMI. (c)-(d) Selected field lines of Model 1 from the high-Q and high twist regions overlaid on $\log Q$ and twist maps, respectively. (e) and (f) present selected field lines from Model 2 overlaid on GST H$\alpha$ and AIA 94~{\AA}~images at 19:38 UT, respectively. (g)-(h) Similar to (c)-(d), but for Model 2.}
\label{fig:fluxropes}
\end{figure}

The model results in comparison with observations are presented in Figure \ref{fig:fluxropes}. Figure \ref{fig:fluxropes}a shows the longitude--latitude map of the radial component of the photopspheric magnetic field (i.e., LOS HMI magnetogram) in the HIRES region of the model at 19:22 UT on 2015 Aug 7. To simulate the observed dark filament threads and flare loops, we insert four flux ropes along the blue curves as shown in Figure \ref{fig:fluxropes}a, in accordance with the flux ropes identified through the NLFFF extrapolations (see Figure~\ref{fig:topology}d). We present two models with sheared arcades and twisted flux ropes. In Model 1, the axial flux of each flux rope is 3$\times10^{20}$ Mx, and the poloidal fluxes are $5\times10^{10}$ Mx cm$^{-1}$ for ropes 1, 4 and  $-5\times10^{10}$ Mx cm$^{-1}$ for ropes 2, 3. The twist of the inserted flux ropes in Model 2 is weaker, i.e., the poloidal fluxes are $1\times10^{10}$ Mx cm$^{-1}$ for ropes 1, 4 and $-1\times10^{10}$ Mx cm$^{-1}$ for ropes 2 and 3,  and the axial fluxes are the same as those in Model 1. The twist angle of the flux rope in the model can be estimated as $\Phi = 2 \pi F_{pol} L / \Phi_{axi}$ \citep{Savcheva2009}. The twist angles in Model 1 after 30000-iteration relaxations is about 4.5$\pi$ for the northern positive ropes and about 2.5$\pi$ for the southern negative one, which is similar to those in the NLFFF extrapolations (see Figure \ref{fig:topology}d2).

Selected field lines traced from the high Q  and high twist regions overlaid on the corresponding $\mathrm{log}\,Q$ and twist maps are presented in the right two columns of Figure \ref{fig:fluxropes}. We can see that for one model the magnetic structure represented with field
lines selected using different methods (i.e., high-Q, high-twist) look slightly different, since the presented field lines are not the same set. For Model 1 the overall structure of the field lines, $\mathrm{log}\,Q$, and twist maps are roughly consistent with the results from NLFFF extrapolations (Figures~\ref{fig:topology}c2--d2). The results in Model 2 with less twist are presented in the bottom row of Figure \ref{fig:fluxropes}. One can see that no coherent flux rope can be identified in Model 2, even if the overall shape of the fields lines appears to match the observed filament threads and flare loops. However, a closer look at the H$\alpha$ images, e.g., Figure~\ref{fig:filament-preview}, suggests that the dark filament threads are highly structured, sometimes display twisted structure, while the field lines in Model 2 is more evenly distributed. Therefore, we think that Model 1 with twisted flux ropes is likely to be a better model for the observations, which is also consistent with the NLFFF extrapolations.

\section{Summary and Discussions}

On 2015 August 7, high-resolution observations of four flares occurring in NOAA AR 12396 have been taken with BBSO/GST. The active region of interest is composed of a leading sunspot group of positive polarity encircled by a chain of following sunspots of negative polarity. We investigated the evolution of various magnetic parameters derived from SDO/HMI magnetograms within one day prior to the studied flares. The evolution of the active region over this time interval is featured by both flux emergence and shearing flows, which is associated with an injection of positive helicity dominated by the shearing flows and also with an injection of positive Poynting flux contributed by both factors. The aforementioned magnetic field observations suggest that this active region was in its emerging phase. The main flaring region encloses bundles of dark filament threads crossing the magnetic polarity inversion line and connecting spots of opposite polarity as shown in the H$\alpha$ observations by GST, conforming to an arch filament system reported in the literature.

To understand the magnetic connectivities within the active region and the evolution, we construct NLFFF models using two independent methods. One method involves extrapolations based on the observed vector magnetograms \citep{Wiegelmann2004}. The semi-circular high-Q line in the east together with the east-west oriented high-Q lines in the west compare favorably with the observed flare ribbons. Magnetic field lines traced along the semi-circular high-$Q$ line are apparently connected to the regions of significant helicity and Poynting flux injection. Based on magnetic twist derived from NLFFF extrapolations, we are able to identify multiple flux ropes consistent with GST observations of filament bundles. Positive twist is dominant in the northern flux ropes, while the southern filament bundles are associated with negative twist. Flux ropes with mixed twist are also observed in NLFFF extrapolations of a twisted jet by \citet{Schmieder2013}.  For a right-handed (positive) twisted tube its sign of helicity is positive, while for a left-handed twisted tube it is negative \citep{Linton2001}. It has been suggested that flux ropes of opposite helicity can interact with each other through magnetic reconnection \citep{Linton2001,Linton2006}. The maps of magnetic twist shows that positive twist became dominant as time progressed, which is consistent with the injection of positive helicity during the same time interval. The other method involves inserting flux ropes into a potential field model constructed based on a LOS HMI magnetogram; the best-fit model is identified by comparison with chromospheric and coronal observations \citep{vanBallegooijen2004}. Using this method, we construct models with sheared arcades and twisted flux ropes. We find that the model with twisted flux ropes matches the observations better, which is also consistent with the NLFFF extrapolations. Our magnetic field models are able to reproduce the magnetic configurations for the observed arch-like H$\alpha$ features. Some of the H$\alpha$ features (i.e., F1/2, F4) appear to be the PIL filaments supported by non-potential magnetic structure \citep{Foukal1971, Gaizauskas1998} rather than the standard arch filaments as shown in \citet{Malherbe1998, Mandrini2002}.

For all of the constructed magnetic field models no relevant null is found. Therefore, although the flares under investigation are characterized by quasi circular-shaped flare ribbons, their occurrence is governed by the AFS characteristics cited above, rather than by a fan-and-spine topology. Flares 1, 2, 4 show similar morphology, and flare brightenings firstly appear at the western border of the AFS with persistent flux cancellations. H$\alpha$ observations by GST show the merge of two short dark filament threads in the AFS into one straightened ``S" shape, which has been reported by \citet{Mandrini2002}. This merging process may lead to the onset of Flare 1, since it is closely related to the timing and location of Flare 1. The onset of Flares 2 and 4 might be caused by magnetic reconnections at the QSLs.

Flare 3 is the largest flare (C5.4) among the sequential flares, and it also behaves differently from others. It begins with motions and brightenings at the two ends of filament F3. The flare brightenings then spread towards the two ends of the northern filament F1/2 and surroundings. During Flare 3, filament F3 appears to evolve from a weakly twisted structure to standard arch filaments after a small rotating motion. Recent high-resolution observations from Hi-C have claimed the first direct evidence of energy release in braided magnetic fields in the corona \citep{Cirtain2013}. Using MHD simulations and forward modeling tools, \citet{Pontin2017} demonstrate that the presence of braided magnetic field lines does not guarantee a braided appearance at the observed intensities. Observed intensities may--but need not necessarily--reveal the underlying braided nature of the magnetic field. Note that currently both observational and modeling studies are focusing on the coronal loops. This is different from our observations of filament threads which sometimes are supported by non-potential magnetic structures. Moreover, the magnetic field modeling also suggests that the underlying magnetic fields are twisted flux ropes. On Aug 7, filaments F1/2 and F3 appear to correspond to the flux ropes with positive twist and negative twist, respectively. However, positive twist becomes dominant in the overall region on Aug 8. The untwisting of filament F3 appears to be consistent with the reduction of negative twist derived from the NLFFF extrapolations. Flare ribbons during Flare 3 match well with the high-Q lines of extrapolated NLFFF model. Flare 3 is similar to the X flare studied by \citet{Inoue2012} in the way that both flares occurred in a region with mixed twist. These results suggest that Flare 3 might be initiated with the magnetic reconnection between the pre-existing F3 and a newly emerging flux rope with positive twist below, which is consistent with \citet{Zuccarello2008}.

A particularly interesting feature that we have identified is the double-ribbon structure on the eastern border of the AFS in Flares 1, 2, and 5. Flare 5 is observed by IRIS on 2015 Aug 8 with slit-jaw image mode. This double ribbon structure is very clear in the high-resolution H$\alpha$ and IRIS observations, but not quite visible in AIA images. One possibility is that the double ribbons indicate the inner/outer edge of a single flare ribbon. This may explain the splitting of ribbon R1 in Flare 5. However, the correspondence between the double ribbon structure and the high-Q lines and the multiple post-flare loop systems (long and short) appears to be in favor of another mechanism, i.e., the double ribbons resulted from two simultaneous reconnections at nearby reconnection sites in the corona.

In summary, the studied AFS contains both standard arch filaments and PIL filaments, i.e., filament bundles supported by twisted flux ropes with mixed helicity (positive in the north and negative in the south) as suggested by the NLFFF modeling. The flare ribbons are consistent with the morphology of QSL footprints in the extrapolated NLFFF, suggesting that all of the flares are associated with magnetic reconnections at the high Q regions. The interesting double-ribbon feature is likely to be caused by simultaneous reconnections at two nearby reconnection sites. The untwisting of filament F3 during Flare 3 appears to be consistent with the reduction of negative twist derived from the NLFFF extrapolations.The high-resolution observations by GST provide us unprecedented details on the fine structure and dynamics of an arch filament system producing multiple flares, thus improving our understanding on the nature of flare initiation and dynamics associated with the AFS.

Acknowledgments: We are grateful to the referee for constructive comments to improve the presentation of the paper. Y.S. acknowledges Drs. Haimin Wang, Yan Xu, Vasyl Yurchyshyn, and Zhi Xu for valuable comments and discussions. This work is supported by NSFC 11473071, 11333009, and U1731241, the Natural Science Foundation of Jiangsu Province (Grants No. BK20141043), and the One Hundred Talent Program of CAS. This work is also supported by the Strategic Priority Research Program on Space Science, the CAS, Grant No. XDA 15052200 and XDA15320301.  R.L. acknowledges the support by NSFC 41774150, 41474151, and 41761134088. We thank the team of BBSO/GST, SDO/AIA, SDO/HMI, and IRIS for providing the valuable data. The HMI data are downloaded via the Virtual Solar Observatory and the Joint Science Operations Center. The BBSO operation is supported by NJIT, US NSF AGS-1250818, and NASA NNX13AG14G grants, and the GST operation is partly supported by the Korea Astronomy and Space Science Institute and Seoul National University and by the strategic priority research program of CAS with Grant No. XDB09000000.

\newpage

\begin{deluxetable}{rrrrrrr}
\tablecolumns{7}
\tablewidth{0pc}
\tablecaption{Timing (UT) and locations of four flares on Aug 7 and one flare on Aug 8 in 2015.}
\tablehead{
\colhead{} & \colhead{Start}   & \colhead{Peak}  & \colhead{Stop} &
\colhead{GOES}    & \colhead{Derived} &\\
\colhead{Flare} & \colhead{time}   & \colhead{time}  & \colhead{time} &
\colhead{class}    & \colhead{position} & \colhead{Instrument}}
\startdata
1 SOL2015-08-07T18:45 & 18:40 & 18:45 & 18:51 & C1.6 & S17E06 & GST, SDO \\
2 SOL2015-08-07T19:10 & 18:53 & 19:10 & 19:24 & C3.4 & S17E08 & GST, SDO \\
3 SOL2015-08-07T19:41 & 19:34 & 19:41 & 19:50 & C5.4 & S17E08 & GST, SDO \\
4 SOL2015-08-07T22:43 & 22:36 & 22:43 & 23:00 & C1.7 & S17E03 & GST, SDO \\
\tableline
5 SOL2015-08-08T02:52 & 02:38 & 02:52 & 03:23 & C1.4 & S18E04 & IRIS, SDO \\
\enddata
\end{deluxetable}
 
\end{document}